\documentclass[a4paper]{easychair}

\usepackage{doc}
\usepackage{makeidx}
\usepackage{color,url}

\hypersetup{colorlinks,pdfborder={0 0 0},urlcolor=black}

\begin{document}

%
\title{Airborne Ultrasonic Tactile Display BCI}


\titlerunning{Airborne Ultrasonic Tactile Display BCI}

%
\author{
Katsuhiko Hamada\inst{1}\thanks{K. Hamada is currently with DENSO Corporation, Japan.}
\and
   Hiromu Mori\inst{2,}\thanks{The author was supported in part by the Strategic Information and Communications R\&D Promotion Program no. 121803027 of The Ministry of Internal Affairs and Communication in Japan.}
\and
    Hiroyuki Shinoda\inst{1}\\
\and
   Tomasz M. Rutkowski\inst{2,3,\dag,}\thanks{The corresponding author.}\\
}

\institute{
  The University of Tokyo,
  Tokyo, Japan\\
\and
   Life Science Center of TARA, University of Tsukuba,
   Tsukuba, Japan\\
\and
RIKEN Brain Science Institute, Wako-shi, Japan\\
\email{tomek@bci-lab.info} \& \url{http://bci-lab.info/}\\
 }

\authorrunning{Hamada, Mori, Shinoda and Rutkowski}

\clearpage

\maketitle

\begin{abstract}
This chapter presents results of our project, which studied whether contactless and airborne ultrasonic tactile display (AUTD) stimuli delivered to a user's palms could serve as a platform for a brain computer interface (BCI) paradigm. We used six palm positions to evoke combined somatosensory brain responses to implement a novel contactless tactile BCI. This achievement was awarded the top prize in the Annual BCI Research Award 2014 competition. This chapter also presents a comparison with a classical attached vibrotactile transducer-based BCI paradigm. Experiment results from subjects performing online experiments validate the novel BCI paradigm.
\end{abstract}


%
%

\pagestyle{empty}

\section{Introduction}
\label{sect:introduction}

State--of--the--art brain computer interfaces (BCIs) are usually based on mental, visual or auditory paradigms, as well as body movement imagery paradigms, which require extensive user training and good eyesight or hearing. In recent years, alternative solutions have been proposed to make use of the tactile modality~\cite{sssrBCI2006,tactileBCIwaiste2010,HBCIscis2012hiromuANDtomek} to enhance  BCI efficiency. 
The concept reported in this chapter further extends the brain's somatosensory channel by applying a contactless stimulus generated with an airborne ultrasonic tactile display (AUTD)~\cite{autdFIRST2008}. This is an expanded version of a conference paper published by the authors~\cite{autdBCIconf2014}.

The rationale behind the use of the AUTD is that, due to its contactless nature, it allows for a more hygienic application, avoiding the occurrence of skin ulcers (bedsores) in patients in a locked--in state (LIS). The AUTD permits a less complex application of the BCI for the caregivers comparing to  classical attached vibrotactile transducers' setups.

This chapter reports very encouraging results with AUTD--based BCI (autdBCI) compared to the classical paradigm using vibrotactile transducer--based oddball (P300 response--based) somatosensory stimulus (vtBCI) attached to the user's palms~\cite{HBCIscis2012hiromuANDtomek}.

The rest of the chapter is organized as follows. The next section introduces the methods used in the study.  The results obtained in online experiments with $13$ healthy BCI users are then discussed. Finally, conclusions are drawn and directions for future research are outlined.

\section{Methods}

Thirteen male volunteer BCI users participated in the reported in this chapter experiments. The users' mean age was $28.54$, with a standard deviation of $7.96$ years. The experiments were performed at the Life Science Center of TARA, University of Tsukuba, at the University of Tokyo and at RIKEN Brain Science Institute, Japan.
The online (real-time) EEG autdBCI and vtBCI paradigm experiments were conducted in accordance with the \emph{WMA Declaration of Helsinki - Ethical Principles for Medical Research Involving Human Subjects} and the procedures were approved and designed in agreement with the ethical committee guidelines of the Faculty of Engineering, Information and Systems at University of Tsukuba, Japan (experimental permission~$2013R7$).

\begin{figure}
	\begin{centering}
	\includegraphics[width=0.7\textwidth,clip]{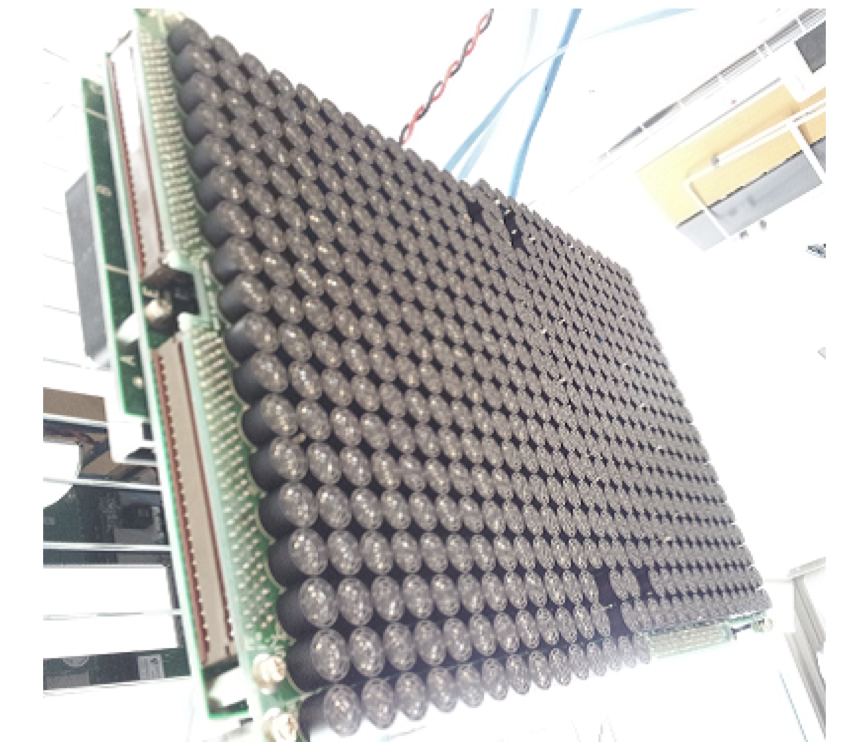}
	\caption{The AUTD array with ultrasonic transducers used to create the
contactless tactile pressure sensation.}
	\label{fig:expFIGautd}
	\end{centering}
\end{figure}

\begin{figure}
	\begin{centering}
	\includegraphics[width=0.7\textwidth,clip]{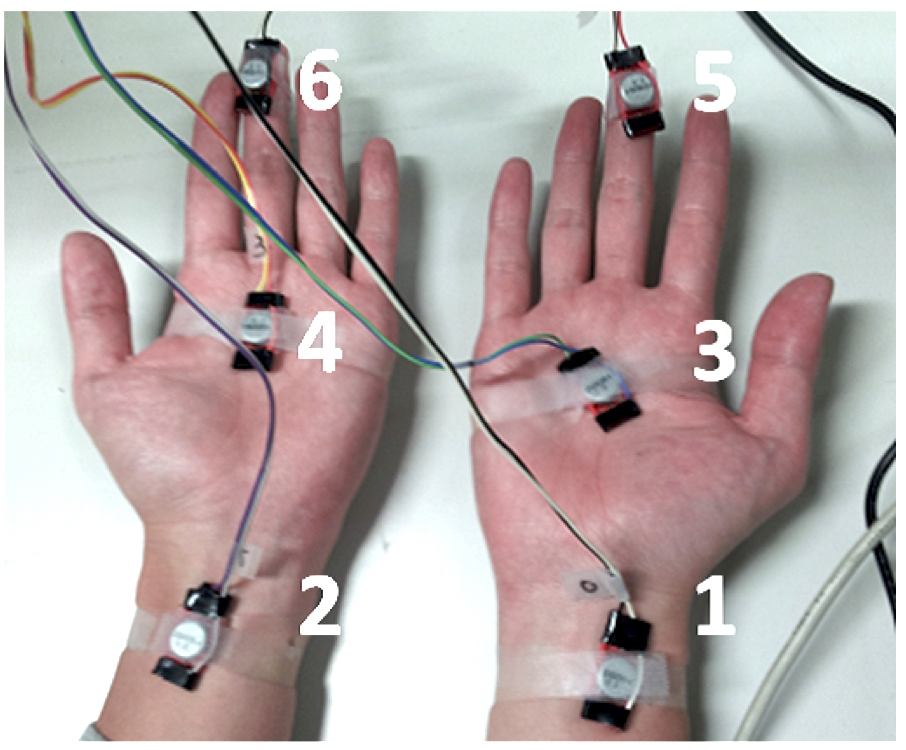}
	\caption{User's palms with attached vibrotactile transducers used in vtBCI experiments. Each stimulus location reflects a different digit.}
	\label{fig:expFIGvtBCI}
	\end{centering}
\end{figure}

The AUTD stimulus generator produced vibrotactile contactless stimulation of the human skin via the air using focused ultrasound ~\cite{autdFIRST2008,katsuhikoAUTD2014}.
The effect was achieved by generating an ultrasonic radiation static force produced by intense sound pressure amplitude (a nonlinear acoustic phenomenon).
The radiation pressure deformed the surface of the skin on the palms, creating a virtual touch sensation.
An array of ultrasonic transducers mounted on the AUTD (see Figure~\ref{fig:expFIGautd}) created the focused radiation pressure at an arbitrary focal point by choosing a phase shift of each transducer appropriately (the so--called phased array technique).
Modulated radiation pressure created a sensation of tactile vibration similar to the one delivered by classical vibrotactile transducers attached to the user's palms, as shown in Figure~\ref{fig:expFIGvtBCI}.
The AUTD device developed by the authors~\cite{autdFIRST2008,katsuhikoAUTD2014} (see Figure~\ref{fig:expFIGautd}) adhered to ultrasonic medical standards and did not exceed the permitted skin absorption levels (approximately $40$ times below the permitted limits).
The effective vibrotactile sensation was set to $50$~Hz~\cite{katsuhikoAUTD2014} to match with tactile skin mechanoreceptors' frequency characteristics and the notch filter that EEG amplifiers use for power line interference rejection.

As a reference, in the second vtBCI experiment, contact vibrotactile stimuli were also applied to locations on the users' palms via the transducers HIHX09C005-8. Each transducer in the experiments was set to emit a square acoustic frequency wave at $50$~Hz, which was delivered from the ARDUINO micro--controller board with a custom battery--driven and isolated power amplifier and software developed in--house and managed from a \emph{MAX~6} visual programming environment.

\begin{figure}
	\begin{centering}
	\includegraphics[width=0.8\textwidth,clip]{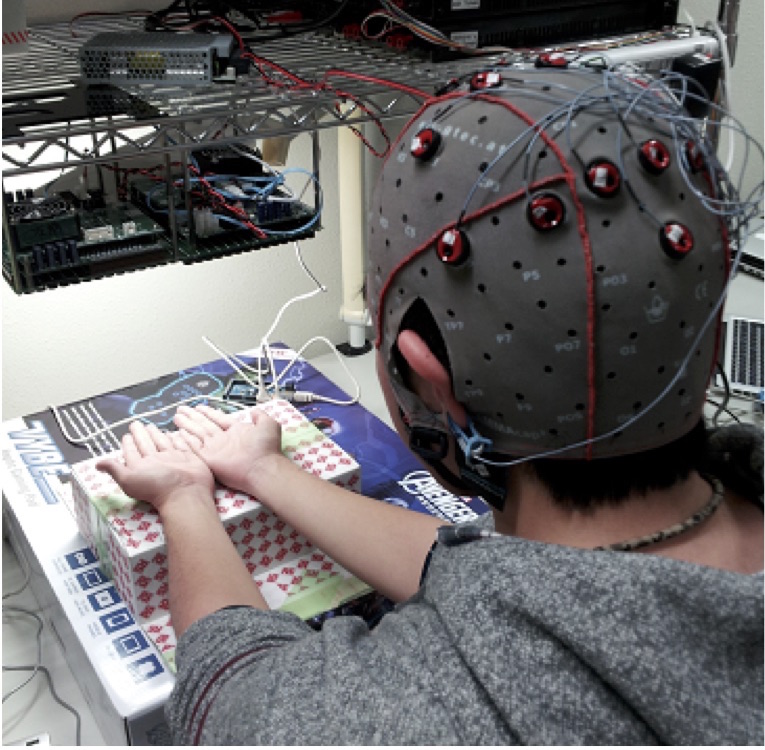}
	\caption{A user during the autdBCI
experiment with both palms
placed under the AUTD array
with ultrasonic transducers.}
	\label{fig:expFIGautdBCI}
	\end{centering}
\end{figure}

The two experiment set-ups above are presented in Figures~\ref{fig:expFIGvtBCI}~and~\ref{fig:expFIGautdBCI}.
Two types of experiments were performed with the volunteer healthy users. Experiments with the target paralyzed users are planned as a followup of the current pilot project. Psychophysical experiments with foot--button--press responses were conducted to test uniform stimulus difficulty levels from response accuracy and time measurements. The subsequent tactile oddball online BCI EEG experiments evaluated the autdBCI paradigm efficiency and allowed for a comparison with the classical skin contact--based vtBCI reference.
In both the above experiment protocols, the users were instructed to spell sequences of six digits representing the stimulated positions on their palms.
The training instructions were presented visually by means of the \emph{BCI2000}~\cite{bci2000book} and \emph{MAX~6} programs with the numbers $1-6$ representing the palm locations as depicted in Figure~\ref{fig:expFIGvtBCI}.

The EEG signals were recorded with the g.USBamp amplifier system from g.tec Medical Engineering GmbH, Austria, using $16$ active g.LADYbird electrodes. The electrodes were attached to the head locations: \emph{Cz, Pz, P3, P4, C3, C4, CP5, CP6, P1, P2, POz, C1, C2, FC1, FC2,} and \emph{FCz}, as in the $10/10$ extended international system. The ground electrode was attached to the \emph{FPz} position, and the reference was attached to the left earlobe. No electromagnetic interference was observed from the AUTD or vibrotactile transducers operating with frequencies notch--filtered together with power line interference from the EEG.
The EEG signals captured were processed online with an in--house extended BCI2000--based application~\cite{bci2000book}, using a stepwise linear discriminant analysis (SWLDA) classifier~\cite{krusienski2006} with features drawn from $0-800$~ms ERP intervals decimated by a factor of $20$.

The stimulus length and inter--stimulus--interval were set to $400$~ms, and the number of epochs to average was set to $15$. The EEG recording sampling rate was set at $512$~Hz, and the high and low pass filters were set at $0.1$~Hz and $60$~Hz, respectively. The notch filter to remove power line interference removed activity between $48 \sim 52$~Hz.
Each user performed three experiment runs (randomized $90$~targets and $450$~non-targets each). As feedback, the spelled numbers (palm position assigned digits as in Figure~\ref{fig:expFIGvtBCI}) were shown on a display to the user. 

\section{Results}\label{sec:results}

\begin{figure}[t]
	\begin{centering}
	\includegraphics[width=\textwidth,clip]{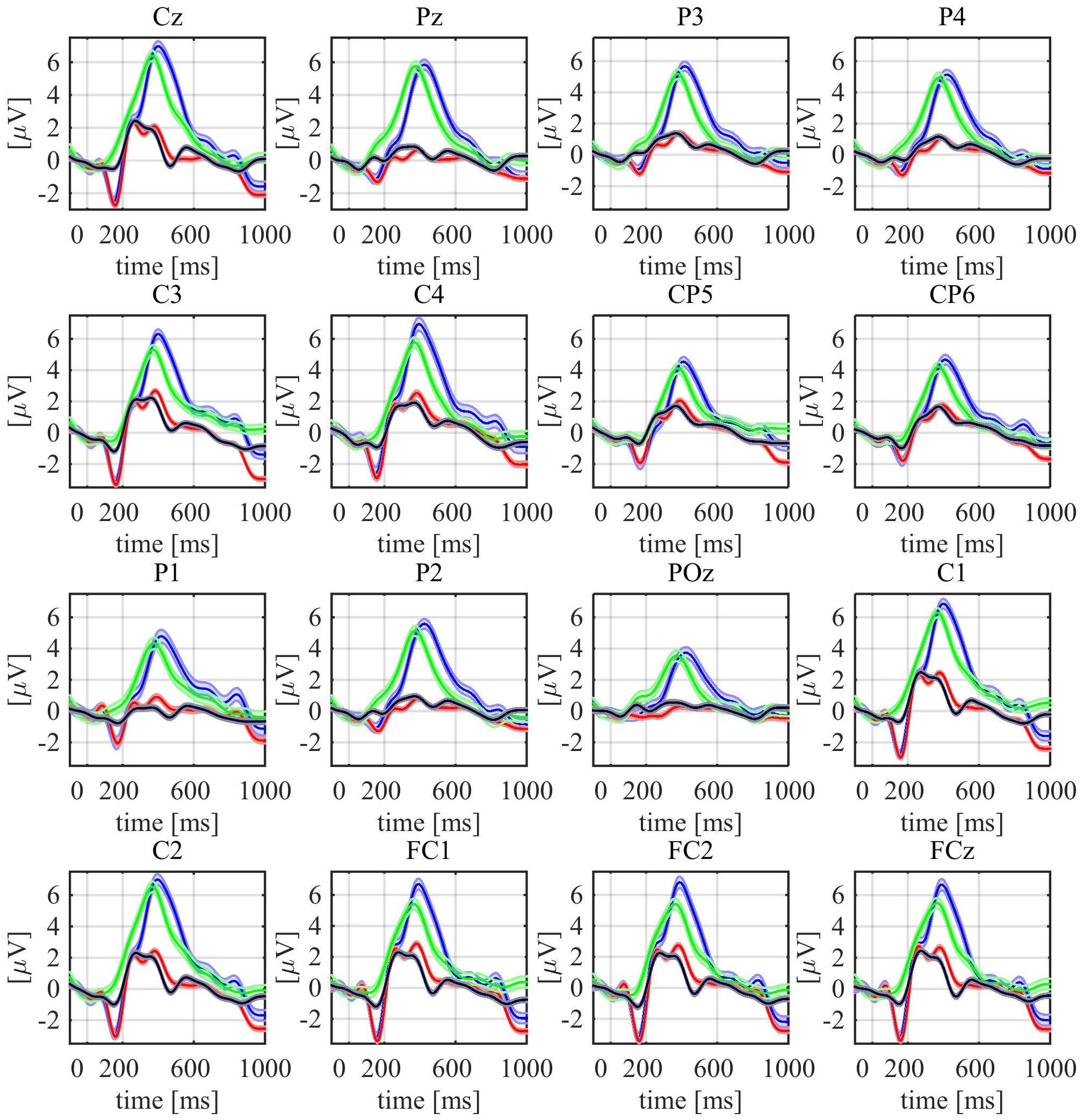}
	\caption{The autdBCI (blue - targets; red - non--targets) and vtBCI (green - targets; black - non--targets) grand mean averaged ERP responses, together with standard error bars.}
	\label{fig:erpFIG}
	\end{centering}
\end{figure}

\begin{figure}[t]
	\begin{centering}
	\includegraphics[width=\textwidth,clip]{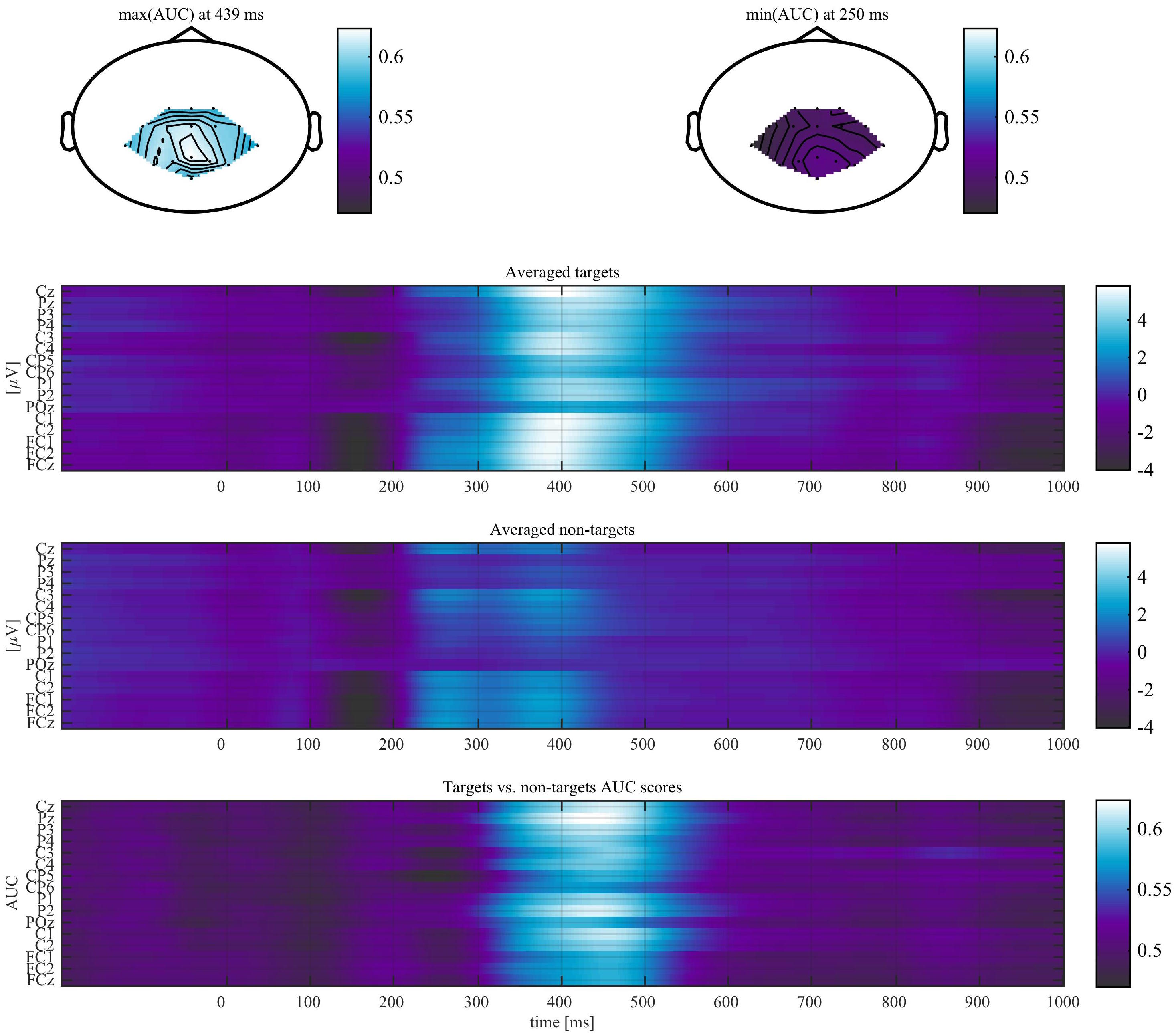}
	\caption{The autdBCI grand mean averaged ERP responses, shown as matrix plots for targets in the top panel; non--targets in the middle; and area under the curve (AUC) of the response discriminability analysis (AUC~$> 0.5$ marks the discriminable latencies). }
	\label{fig:aucFIG}
	\end{centering}
\end{figure}

\begin{figure}[t]
	\begin{centering}
	\includegraphics[width=\textwidth,clip]{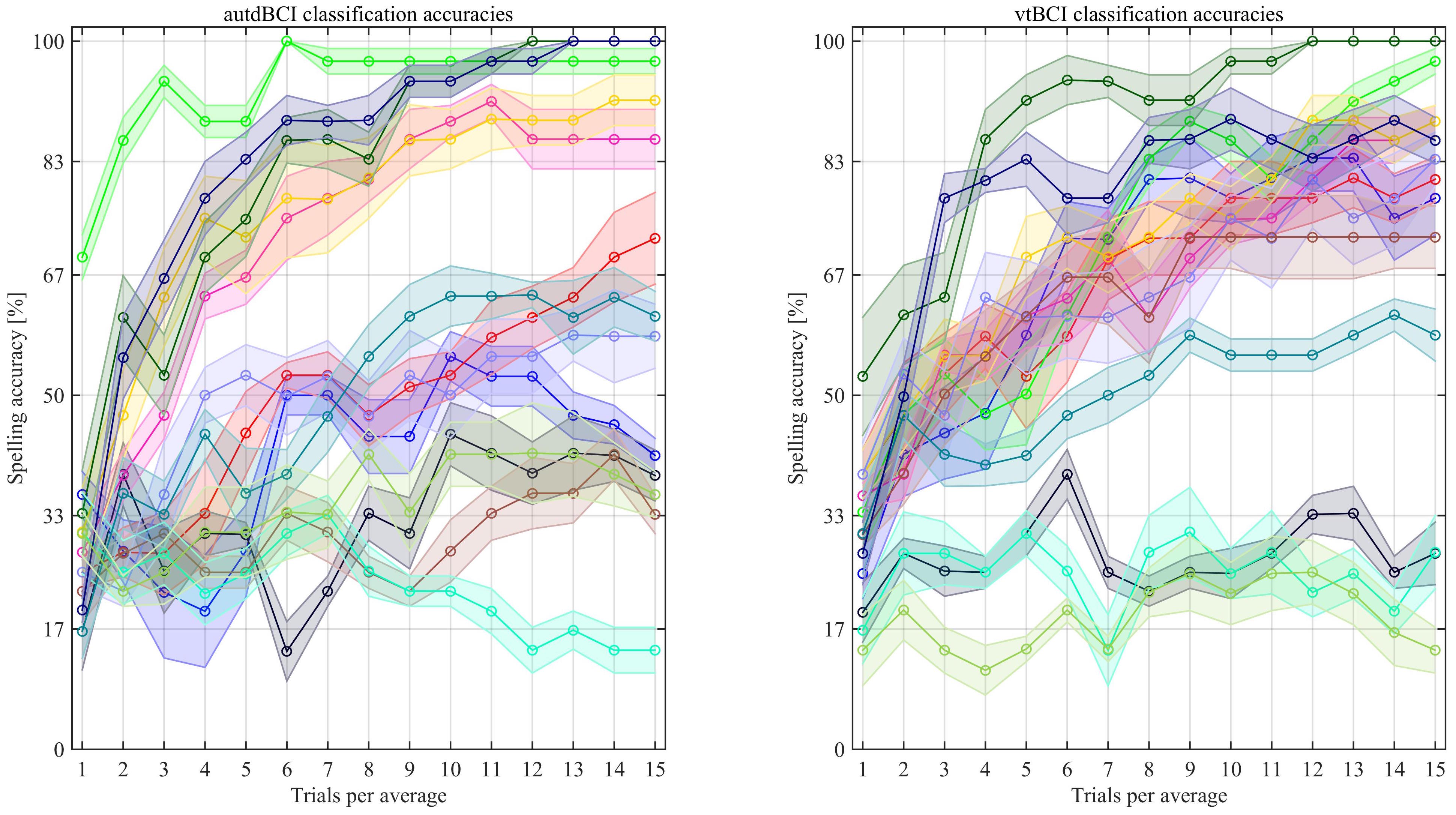}
	\caption{Averaged autdBCI and vtBCI spelling accuracy across six digits, colour coded for each user with standard error bars.}\label{fig:accuracy}
	\end{centering}
\end{figure}

The grand mean averaged evoked responses to targets and non--targets are depicted together with standard error bars in Figure~\ref{fig:erpFIG} and as matrices with an area under the curve (AUC) analysis for feature separability in Figure~\ref{fig:aucFIG}. The BCI six digit sequences spelling accuracy analyses for both experiments for the various averaging options are summarized in Figure~\ref{fig:accuracy}. The chance level was $16.6\%$. The mean six digit sequence spelling accuracies for $15$-trial averaged ERPs were $63.8\%$ and $69.4\%$ for autdBCI and vtBCI, respectively. The maximum accuracies were $78.3\%$ and $84.6\%$ respectively. The differences were not significant, supporting the concept of using autdBCIs. However, a single trial classification offline analysis of the collected responses resulted with the best obtained accuracies of $83.0\%$ for autdBCI and $53.8\%$ for vtBCI, leading to a possible $19.2$~bit/min and $7.9$~bit/min, respectively. 

In the case of the autdBCI, only a single user's results were bordering on the level of chance, and four subjects attained $100\%$ ($10$ trials averaging). On average, lower accuracies were obtained with the classical vtBCI, with which three users bordered on the level of chance, and only one user scored $100\%$ accuracy level in SWLDA--classified averaged responses.

\section{Conclusions}

This study demonstrates results obtained with a novel six--command--based autdBCI paradigm. We compared the results with a classical vibrotactile transducer stimulus--based paradigm. The experiment results obtained in this study confirm the validity of the contactless autdBCI for interactive applications and the possibility to further improve the results through single trial--based SWLDA classification.

The EEG experiment with our paradigm confirms that contactless (airborne) tactile stimuli can be used to create six command--based BCIs in real time. A short demo with online application of the paradigm to robotic arm control is available on YouTube~\cite{youtubeAUTDbciROBOT}.

The results presented offer a step forward in developing and validating novel neurotechnology applications. Since most users did not achieve very high accuracy during online BCI operation, especially with only a few trials, the current paradigm obviously requires improvement and modification. These requirements determine the major lines of study for future research. 

However, even in its current form, the proposed autdBCI can be regarded as a practical solution for LIS patients (locked into their own bodies despite often intact cognitive functioning), who cannot use vision or auditory-based interfaces due to sensory or other disabilities. The reported autdBCI project was awarded The BCI Annual Research Award 2014 for ``A fascinating new idea never explored before,'' according to the Chairman of the Jury for the Annual BCI Research Award 2014, Prof.~Gernot~R. Mueller-Putz from the Institute for Knowledge Discovery, Graz University of Technology, Austria.

\end{document}